\documentclass[showpacs,amsmath,amssymb,prb,preprint]{revtex4}

\usepackage{graphicx}% Include figure files
\usepackage{booktabs}
\usepackage{textcomp}

\begin{document}

\title{ Characteristics of phonon transmission across epitaxial interfaces: a lattice dynamic study  }

\author{Jian Wang}
\author{Jian-Sheng Wang}

\affiliation{Center for Computational Science and Engineering and
Department of Physics, National University of Singapore, Singapore
117542, Republic of Singapore}

\date{28 March 2007}
%\date{\today}

\begin{abstract}
  Phonon transmission across  epitaxial interfaces is studied
  within the lattice dynamic approach. The transmission shows weak dependence on frequency for the lattice wave with a fixed angle of incidence.
  The dependence on azimuth angle is found to be related to the symmetry of the boundary interface.
   The transmission varies smoothly  with the change of the incident
  angle.  A critical angle of incidence exists when the phonon is incident from  the side with  large group
  velocities to the side with low ones. No significant mode conversion is observed among different acoustic wave branches at the interface, except when the incident angle is
near the critical value.  Our theoretical result of the Kapitza
conductance $G_{K}$ across the Si-Ge (100) interface at temperature
$T=200\;$K is $4.6\times10^{8}\; \mbox{\rm WK}^{-1}\mbox{\rm
m}^{-2}$. A scaling law $G_K \propto T^{2.87}$ at low temperature is
also reported. Based on the features of transmission obtained within
lattice dynamic approach, we propose a
 simplified formula for  thermal conductance
across the epitaxial interface. A reasonable consistency is found
between the calculated values and the experimentally measured ones.

\end{abstract}


% from PRB 72.10.Bg, 05.60.Gg, 63.22.+m, 66.70.+f

\maketitle

\section{Introduction}
 The interfacial thermal conductance plays a critical role in
nanometer-scaled devices\cite{dgcahill,ETswartz}.  Kapitza
resistance \cite{Kapitza} was first discovered in 1940s. Although
much work, both theoretical and experimental, \cite{dgcahill,
ETswartz,Kapitza, walittle, G-Chen, experimental,dayang, zhao, wang}
has been done since then, the characteristic behavior of phonon
transmission across the interface is still not clear. The continuum
elastic wave model \cite{walittle} may be inaccurate when the
details of atomistic structures and phonon dispersions are
considered. The diffusive scattering theory\cite{ETswartz} neglects
the wave property of phonons. Here we study phonon transmission
across interfaces with the lattice dynamic
approach\cite{dayang,wang}, which  can simulate the phonon
transmission across interfaces atomistically from the
first-principles. We concentrate on the phonon transmission at a
single epitaxial interface to capture salient features of
transmission.

The paper is organized as follows. We first present the method
employed in this study. Using this method we calculate the
dependence of phonon transmission on frequency, the azimuth angle,
the incident angle, the mode conversion at the Si-Ge interface and
at the Si-GaP interface in Section II. Some features of phonon
transmission are obtained and discussed. Finally, on the basis of
these features, we propose a simplified formula for thermal
conductance across interface. A comparison  between the calculated
thermal conductance and the experimentally measured values is made.

\section{ Method }

We consider the problem of phonon transmission across an epitaxial
interface. Two types of important crystal structures in
semiconductors, diamond and zinc-blend, are chosen in this study.
The crystals on each side of the interface are assumed
semi-infinite\cite{dayang,wang,zhao}. For simplicity, we also
neglect the defects possibly existing between the two crystal solids
and assume that the interface is an ideal epitaxial interface.
Lattice dynamic approach also requires that the harmonic
approximation for solids is valid. This condition is usually
satisfied at low temperatures. For the scattering of waves at
interface, we remark that the nonlinear nonlinear effect may be less
important than the elastic scattering at  low temperature because
the thickness of boundary area at the interface is smaller than the
mean free path of phonons.

We follow the  scattering boundary method in
Ref.~\onlinecite{dayang, wang}. We write the solutions on the
incident side and on the transmitted side with the unknown component
coefficients  expressed as follows.  If a normal mode  $ \tilde {\bf
u}^{L}_{l,i,n}(\omega, {\bf q})$
 is incident from the left lead,
the scattering solution for the perfect leads  can be assumed as
\begin{subequations}
\label{boudaryeq}
\begin{eqnarray}
{\bf u}_{l,i}^{L} &= & \tilde {\bf u}_{l,i,n}^{L}(\omega, {\bf q}_n) + \sum_{n'}t^{LL}_{n'n} \tilde {\bf u}_{l,i,n'}^{L}(\omega, {\bf q}'_{n'}), \\
{\bf u}_{l,i}^{R}&=& \sum_{n''}t^{RL}_{n''n} \tilde {\bf
u}_{l,i,n''}^{R}(\omega, {\bf q}''_{n''}),
\end{eqnarray}
\end{subequations}
where $i,l$ denote the $i$th atom in the $l$ unit cell, and $n, n'$
and $ n''$ refer to  the different polarized branches of incident,
reflected and transmitted  waves. Frequency is denoted as $\omega$.
Wavevectors  for the incident, reflected, and transmitted waves are
${\bf q}, {\bf q}', $ and ${\bf q}''$, respectively.  The
superscript $L$ and $R$ indicate the left and the right.
 In these equations, $t^{RL}_{n''n}$, $t^{LL}_{n'n}$ are
the amplitude transmission/reflection coefficients from mode $n$ on
the lead $ L $ to mode $n''$ on lead $R$, and to mode $n'$ on lead
$L$. The wave vectors $\bf q'$ and ${\bf q}''$  satisfy
$\omega=\omega_{n'}({\bf q'}) = \omega_{n''}({\bf q}'')$. Note that
frequency does not change because the system is linear.

In this paper, we consider the phonon transmission across two types
of epitaxial interfaces: the interface between silicon and germanium
and the interface between silicon and  gallium phosphide. Now the
problem is to determine the wave numbers for the possible branches
of the reflected and the transmitted waves.  Since the system is
homogeneous in the $x$ and $y$ directions, the transverse components
of the wave vector for the reflected waves ${\bf q}'$ and  for the
transmitted waves ${\bf q}''$ have the same values as that of the
incident wave ${\bf q}$, that is $q'_{x}= q''_{x}=q_{x}$;
$q'_{y}=q''_{y}=q_{y}$.  The longitudinal components,  $q'_{z}$ and
$q''_{z}$, satisfy $\omega({\bf q})=\omega({\bf q'})=\omega({\bf
q''})$.    We solve the nonlinear equation to find $q'_z$ or $q''_z$
for the $(111)$ interface with the numerical method  in
Ref.~\onlinecite{numericalrecipes}.   For the $(100)$ interface, we
utilize a more efficient eigenvalue method to obtain the solutions
of nonlinear equations.\cite{zhao}  The wave vectors $q'_z$ and
$q''_z$
 are identified from these solutions to satisfy the following conditions: (i) The
reflected waves should have the negative group velocity $v_z({\bf
q}')$ so that the phonon energy of the reflected waves will
propagate back to the $-z$ direction. (ii) The transmitted wave
should have the positive group velocity so the phonon energy of the
transmitted waves will propagate to the $+z$ direction. (iii)When
the wave numbers for the reflected and transmitted waves are
complex, they are identified from the solutions to satisfy  ${\bf
Im}(q'_z)<0$ and ${\bf Im}(q''_z)>0$ so that these waves  decay in
space. Actually these decay waves with complex wave number do not
propagate any energy and the phase difference along $z$ direction
for atoms in the neighboring unit cells is $\pi$.  But they are
physically possible  modes at the interface and must be considered
in the scattering boundary method.

\begin{figure}[bht]
\includegraphics[width=0.9\columnwidth]{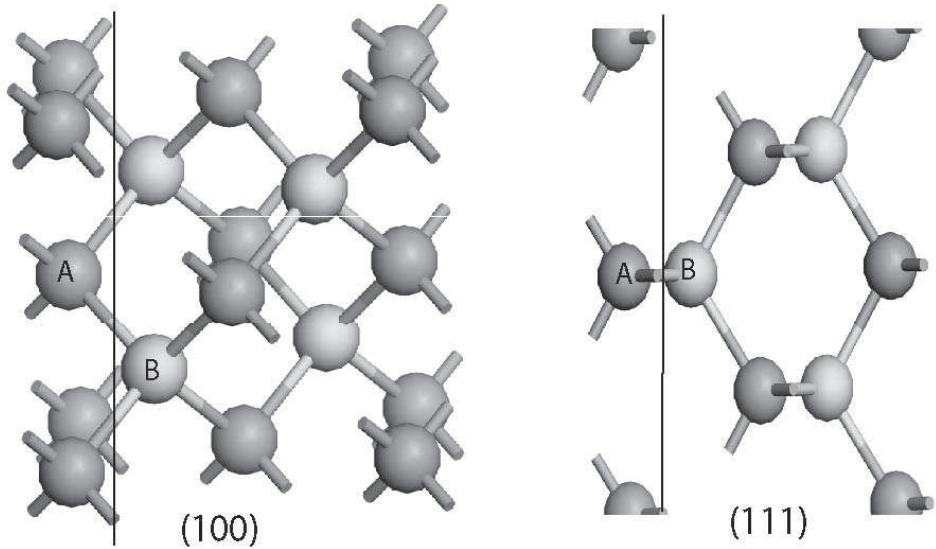}
\caption{ \label{fig:interface} Atomic interfacial structure of the
(100) and (111) interface.}
\end{figure}

Now we consider the dynamic equations for atoms at the interface.
The $(100)$ and $(111)$ interface between the diamond structure or
the zinc-blend structure is illustrated in Fig.~\ref{fig:interface}.
We assume that  the transverse directions at the interface are
infinitely large so that the boundary atoms on each side, e.g., $A$
atoms in the figure, are   translational invariant. The dynamic
equations for $A$ position  atoms are equivalent, and similarly for
$B$ position atoms on the other side of the interface.  So we only
need to consider the dynamic equations for atom $A$ and for atom $B$
at the interface. Under the harmonic approximation, these dynamic
equations at a given frequency $\omega$ can be written as $ -m_i
\omega^2 {{\bf u}}_{i} + \sum\limits_{j}{\bf K}_{i,j} \cdot {\bf
u}_{j}=0, $ where $m_i$ denotes the atomic mass for each atom,
${{\bf u}}_{i}$ is the oscillation amplitude and ${\bf K}_{i,j}$ is
the force constants. The force constants across the  interface are
chosen as the symmetrized $ K_{ij}=(K_{ij}^{L}+K_{ij}^{R})/2 $.
After substituting Eq.(\ref{boudaryeq}) into this dynamic equation,
we get six equations with six unknown coefficients. This system of
equations
 can be solved by a conventional method.   The energy
transmission from mode $(L , n)$ to
 mode $ (R, n'')$ is given by
\begin{equation}
  \label{energytran}
 \tilde{T}_{n''n}^{RL}= |t_{n''n}^{RL}|^{2}\frac{\tilde{v}^{R}_{n''}}{\tilde{v}^{L}_{n}}.
\end{equation}
Here $\tilde{v}^{L}_{n}$ is the  reduced group velocity \cite{wang}
along $z$ direction for mode $n$ in the left lead, defined by
$\tilde{v}^{L}_{n}=v^L_{n}/l_L$ where $v^L_{n}$ is the group
velocity and $l_L$ is the lattice constant for the left. Similar
meaning holds for $\tilde{v}^R_{n}$ . The total reflection
$\mathcal{R}_{n}^{L}$ and transmission coefficients
$\mathcal{T}_{n}^{L}$ for $(L , n) $ are given, respectively,  by
\begin{subequations}
\label{trantotal}
\begin{eqnarray}
 \label{reftran}
 \mathcal{R}_{n}^{L}= \sum _{n'}|t_{n'n}^{LL}|^{2}\frac{ \tilde{v}^{L}_{n'}}{ \tilde{v}^{L}_{n}},
 &&
 \mathcal{T}_{n}^{L}= \!\!\sum _{n'' }|t_{n'n}^{RL}|^{2}\frac{ \tilde{v}^{R}_{n''}}{ \tilde{v}^{L}_{n}}, \\
\mathcal{R}_{n}^{L} + \mathcal{T}_{n}^{L}  \equiv  1.&&
 \label{reftrann2}
\end{eqnarray}
\end{subequations}
  The group velocity along $z$ direction is
calculated through the dynamic matrix as
\begin{equation}
\label{groupvelocity} v_z = \frac{1}{2\, \omega} \frac{{\bf
\tilde{e}}^{\dag} \frac{\partial {\bf D} }{\partial q_z}{\bf
\tilde{e}}, }{{\bf \tilde{e}}^{\dag}\cdot {\bf \tilde{e}} },
\end{equation}
where ${\bf D}$ is the dynamic matrix and ${\bf \tilde{e}}$ the
eigenvector of the dynamic matrix\cite{wang}.  Note that the energy
conservation relation, Eq.(\ref{reftrann2}),  is satisfied
automatically. This can be used as a check. With the relation
Eq.~(\ref{trantotal}), we can get the Kapitza conductance $G_{K}$ as
\begin{equation}
\label{current}
 G_{K}  = \frac{1}{V} \sum \limits_{{\bf q}, n} \hbar
 \omega_{n}({\bf q})v_{n}^z({\bf q}) \mathcal{T}_{n}({\bf q},
 \omega_{n}) \frac{\partial f(\omega_n, T)}{\partial T},
 \end{equation}
where $\mathcal{T}_{n}({\bf q}, \omega_{n})$ is the transmission
\cite{wang} calculated from Eq.~(\ref{trantotal}), $V$ is the volume
and $f(\omega_n, T)$ is the Bose-Einstein distribution for the
$n-{\rm th}$ branch mode. When using Eq.~(\ref{current}), we compute
Kapitza conductance from the Si side. The lattice constants for Si
 conventional unit cell is $a=5.43\,$\AA .

\section{Models and Numerical Results}

In this section, we  report the dependence of phonon transmission on
frequency, the azimuth angle, and the incident angle  across the Si
and Ge interface. We then consider the mode conversion problem at
interface. Phonon transmission across a kind of zinc-blend
interface: Si-GaP is also studied to investigate the lattice
structure effect on transmission.
\begin{figure}[bht]
\includegraphics[width=0.9\columnwidth]{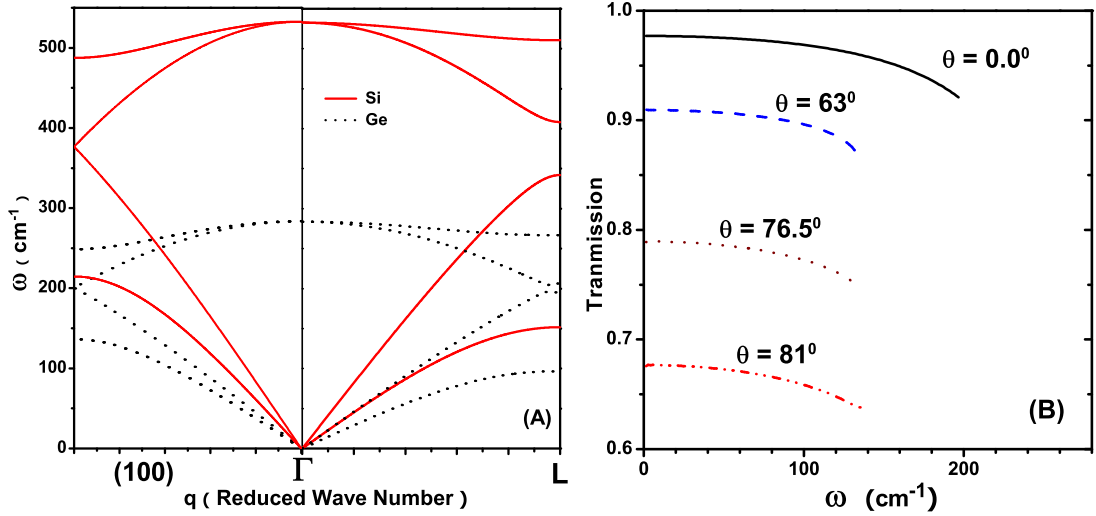}
\caption{ \label{fig:dispersion} (A) Phonon dispersion for Si and Ge
along $\Gamma - \mbox{L}$ and $(100)$ direction. (B) The energy
transmission $T_n(\omega)$ as a function of  frequency $\omega$
across the Si-Ge (100) interface for $LA$ waves at different angles
of incident from Si to Ge.}
\end{figure}
The atomic masses for Si and Ge are $28 \, \mbox{\rm {amu}}$ and $
72.61 \, \mbox{\rm {amu}}$, respectively.  After optimizing the
structure using Tersoff potential \cite{dwbrenner}, we obtain the
linearized force constants under small displacement.  Due to Tersoff
potential truncation function,\cite{dwbrenner} only four nearest
atoms need to be considered for each atom. The phonon dispersions
for the Si and Ge along the $\Gamma-L$ direction can be calculated
through the linearized force constants from the dynamic matrix and
are illustrated in Fig.~\ref{fig:dispersion}. The maximum frequency
of LA branch along $\Gamma-L$  for Si and Ge are $343\, {\rm
cm}^{-1}$ and $ 196\, {\rm cm}^{-1} $, respectively. Compared with
the experimental Si and Ge phonon dispersions \cite{giannozzi},
there is a discrepancy between the calculated phonon frequency in
Fig.~\ref{fig:dispersion} and the experimental values. This
inaccuracy comes from the lack of more neighboring atoms for Tersoff
potential, apart from the first neighboring ones. The \emph{ab
initio} method\cite{abinit} which can include higher order
neighboring atoms is desirable for the more elaborate phonon
dispersion relations. However, we find that the inclusion of higher
order neighboring atoms will make the calculation of energy
transmission \ very difficult because the solutions of $q'_z$ or
$q''_z$ at a given frequency $\omega $ is rather complicated. We
estimate that such errors of phonon dispersion relation in the
Briullouin boundary will not have much influence on the energy
transmission across the interface.  The force constants from the
first neighboring atoms play a main role in the computation of
phonon dispersion relations. The cut-off frequency of energy
transmission across the interface will not approach closely to the
boundaries of the Briullouin boundary apart from a few high symmetry
directions. The transmission decreases very rapidly when the
frequency is near cut-off, which can be seen from Fig.2(B).
Therefore, we think that the inaccuracy of the Briullouin  boundary
values from Tersoff potential will not significantly change the
results of energy transmission across the interface.

\subsection{ Dependence on angular frequency}
The transmission's dependence on angular frequency across a $(100)$
Si-Ge interface is shown in Fig.~\ref{fig:dispersion}(B). The
incident acoustic wave is  longitudinally polarized with an incident
angle $\theta$ in $xz$ plane.  For waves with the incident angle
$\theta$ below about $63^\circ$, the transmission falls into a very
narrow range approximately from $0.91$ to $0.98$ and decreases
slightly with the increase of frequency before it reaches its
cut-off frequency. Here the  cut-off frequency means that beyond
this point the phonon transmission equals zero. There is an  abrupt
decrease in phonon transmission when the frequency is near this
value. When the incident wave is normal to the interface along
$(100)$ direction from Si to Ge, the transmitted waves are also
along $(100)$ direction. Thus, the cut-off frequency of transmission
can be easily determined from the phonon  dispersion of Ge. This can
be shown from Fig.~\ref{fig:dispersion}(B). The cut-off frequency is
about $197 \rm{\mbox{ cm}}^{-1}$  for the wave with the incident
angle $\theta =0$, while the endpoint frequency of the longitudinal
acoustic wave branch for Ge along $(100)$ direction is $200
\rm{\mbox{ cm}}^{-1}$,  which agrees with the cut-off frequency of
transmission for the normal incident wave. In
Fig.~\ref{fig:dispersion}(B), for  clarity, we only plot the
transmission before its abrupt jump to zero.   For transversely
polarized incident acoustic waves, we also have observed a similar
phenomenon. It can be concluded that the acoustic phonon
transmission from Si to Ge varies in a narrow range with the
increase of frequency, independent of the the property of its
polarization. Note that the phonon wave is incident from the
material (Si) with a larger group velocity to the material (Ge) with
a smaller group velocity.  This conclusion also holds in our
calculations of the transmissions for other incident acoustic waves,
not only in $xz$ plane, but also with azimuth angle $\varphi$.  We
remark that this small change of phonon transmission with the
frequency for incident wave at the incident angle $\theta$ can be
neglected for a approximate estimation of thermal conductance across
an interface.

\subsection{Spatial angular dependence}
\subsubsection{Azimuth angle and symmetry of the interface}
\begin{figure}[ht]
\includegraphics[width=0.9\columnwidth]{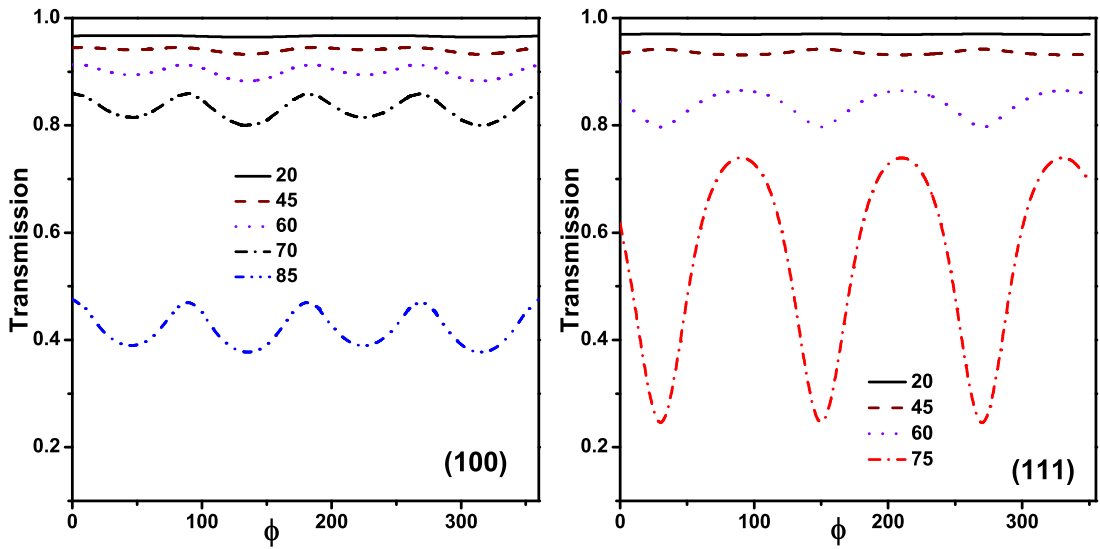}
\caption{ \label{fig:pha} Dependence of the longitudinal acoustic
phonon transmission on the azimuth angle $\varphi$. All
transmissions plotted are at angular frequency $\omega = \mbox{15
cm}^{-1}$ and are incident from Si to Ge. Different lines in the
figure correspond to the transmissions for wave with different
incident angle $\theta$ as the indicated   in the figure. Left
figure shows the results for the $\mbox{(100)}$ interface,
  right figure  for the  $\mbox{(100)}$ interface.  }
\end{figure}

We find that the phonon transmission's dependence on the azimuth
angle of the incident wave is related to the symmetry of the
interface.  The results of the transmissions for the longitudinal
acoustic waves incident from Si to Ge  are shown in
Fig.~\ref{fig:pha}. It can be seen from  Fig.~\ref{fig:pha} that the
phonon transmission shows little dependence on the azimuth angle
 $\varphi$ at incident angle $\theta <45^\circ$ both for the $(100)$ and
$(111)$ interface. However, when the incident angle increases beyond
$45^\circ$, there are  peaks in transmission with the variation of
the azimuth angle. The peaks reveal the anisotropy of the interface.
There is a four-fold symmetry axis in the $[100]$ direction in Si or
Ge and a three-fold symmetry axis along the $[111]$ direction. These
symmetries have evidently been shown in the phonon transmission
across the related interface. In Fig.~\ref{fig:pha}, there are
four-fold symmetrical peaks separated by $\pi/2$  in the
transmission across the $(100)$ interface. Three three-fold
symmetrical peaks separated by $2\pi/3$ appear in the transmission
across the $(111)$ interface.

\subsubsection{Critical incident angle}
\begin{figure}[h]
\includegraphics[width=0.8\columnwidth]{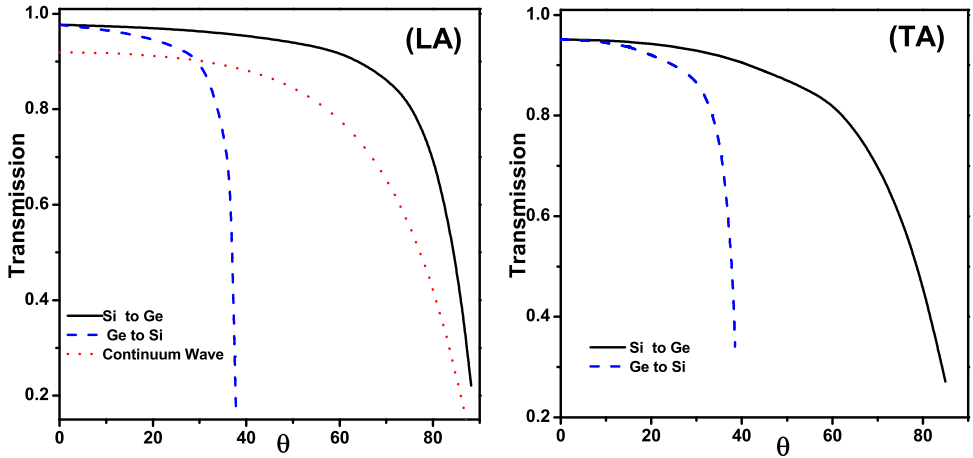}
\caption{ \label{fig:critical_angle} Dependence of phonon
transmission on the incident angle $\theta$. The left figure shows
the results of longitudinal acoustic waves $\mbox{(LA)}$ and the
right figure is for the transverse polarized waves $\mbox{(TA)}$.
The   solid lines indicate the transmission from Si to Ge and the
dashed lines denote the transmission from Ge to Si. All the
transmissions by lattice dynamic approach are calculated at angular
frequency $ \omega = \mbox{15 cm}^{-1} $. The   dotted line in the
left figure is the result calculated from the continuum wave model.
}
\end{figure}

 We next report the result of the dependence of energy
transmission on the incident angle. A continuum wave incident on a
surface with an incident angle $\theta$  is refracted in accordance
with Snell's law. It would be an interesting question if there is a
critical angle for discrete lattice waves.  We calculated the phonon
transmission incident from Si to Ge and incident from Ge to Si. The
dependence of transmission on the incident angle are illustrated in
Fig.~\ref{fig:critical_angle} for the $(100)$ Si-Ge
interface. It
can be seen from Fig.~\ref{fig:critical_angle} that when the angle
$\theta$ for the phonon incident from Si to Ge increases, energy
transmission decreases slowly. The incident angle from Si to Ge can
be extended to  as large as $90^{\circ}$. In contrast to the
transmission from Si to Ge, for waves incident from Ge to Si, there
exists a critical angle, about $38^{\circ}$ for $LA$ and
$38.5^{\circ}$ for $TA$, above which the transmission is very small.
We can estimate the critical angle with the help of Snell's law for
continuum wave. The group velocities calculated from the dynamic
matrix in this paper for longitudinally polarized waves along the
$[001]$ direction are $v_{L}^{Si}\approx 6.87\,$Km/s and $v_{L}^{Ge}
\approx 3.78\,$Km/s, repectively. The Snell's law of continuum wave
model gives the critical angle
 $ \theta_{c}=\sin^{-1}(v_{L}^{Ge}/v_{L}^{Si})=
33.4^{\circ}$ from Ge to Si. This value is a little lower than the
observed critical angle value. However, we can still  think that the
Snell's law holds approximately.

When the incident waves from Si to Ge and from Ge to Si are both
normal to the surface ($\theta = 0 $), the  transmissions from both
sides are $0.98$ for $LA$ waves and $0.95$ for $TA$ waves.  We use
the acoustic mismatch model \cite{walittle} to estimate
$4Z_{Si}Z_{Ge}/(Z_{Si}+Z_{Ge})^{2}\approx0.97 \mbox{ for LA}, 0.94
\mbox{ for TA} $, where $Z_{Si}$ and $Z_{Ge}$ denote the acoustic
impedance defined as $Z=\rho v$. Here $\rho$ is the mass density and
$v$ is the group velocity.  Here we take $\rho _{Si}=2.329\times
10^{3}\mbox{ Km/s}$ and $\rho _{Ge}=5.323\times 10^{3}\mbox{ Km/s}$.
The values of group velocities for $LA$ waves are stated in the
previous paragraph.  It can be seen that the acoustic mismatch model
in Ref~\onlinecite{walittle} well describes the transmission for
normal incident waves.

To investigate the difference in the transmission on the incident
angle between the lattice dynamic approach and the continuum wave
model, we also calculated the energy transmission of the continuum
wave. \cite{auld} The results for Si-Ge interface is illustrated in
Fig.~\ref{fig:critical_angle}. The procedure of calculation is as
follows. First, the amplitudes transmission \cite{auld} is computed
from the Fresnel equation $ t= \frac{2z \cos \theta }{z \cos \theta
+ z' \cos \theta '} $, where $z=\sqrt{\rho c}$ and $z'=\sqrt{\rho '
c'}$ is the wave impedance for the incident and refraction crystal.
Here the mass density $\rho$ and $\rho '$ take the value in the
preceding graph. The stiffness constants $c, c'$ are $0.796, 0.680
\times10^{11}\mbox{ \rm{N/m}}^2$, respectively. The angle of
incidence $\theta$ and the angle of transmission $\theta '$ satisfy
the Snell's Law. The energy transmission for the continuum wave is
calculated through the formula $ \mathcal{T}=|t|^2\frac{\rho ' v'
\cos \theta '}{\rho v \cos \theta }$, where $v =8.43 \mbox{Km/s}$,
$v'=4.87 \mbox{Km/s}$ are the group velocities for the incident and
refraction continuum waves. It can be seen that the isotropic
continuum wave gives a similar dependence of transmission of the
incident angle.  The result for the continuum wave is close to the
result of lattices wave.

\subsection{Mode conversion at the interface}
 The mode conversion at the interface would be an interesting
 problem to pursue. There are two types of
 mode conversion for thermal transmission across the boundary: acoustic-optical (AO) and
 acoustic-acoustic (AA). We neglect the optical-optical conversion
 because the corresponding frequencies do not overlap  in our model. We think that
 this optical-optical conversion's  contribution to thermal transport is trivial because
 of its relatively low group velocity and high energy.

 \emph{AO conversion.} We did not observe  significant AO
 conversion in our simulation of phonon transmission at the Si-Ge
 interface. The AO conversions in our results are very small, no more
 than $10\%$ and fall in a very narrow frequency range. This result
 is different from the reported results in Ref.~\onlinecite{zhao}.  We think this maybe
 come from the empirical potential chosen in our paper, which
 results in different group velocities. To estimate the maximum ratio of  a possible acoustic-optical mode
conversion, A simplified 1D toy chain model \cite{wang} can be
composed because it can be assumed that the AO conversion will be
easier for both longitudinal waves in one dimension. However, the
results in Ref.~\onlinecite{wang} find that the energy transmission
contributed by the AO conversion is trivial in comparison with the
acoustic-acoustic transmission. This is due to the large mismatch in
their group velocities.

\begin{figure}[hbt]
\includegraphics[width=0.8\columnwidth]{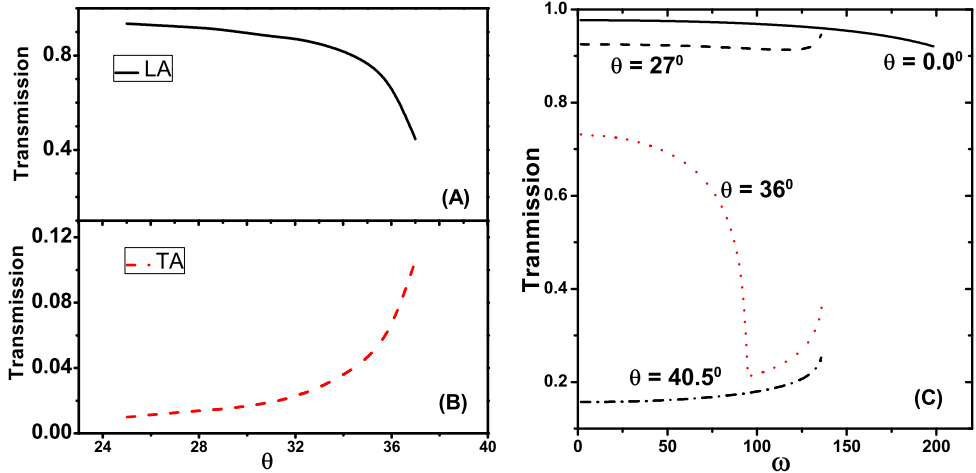}
\caption{ \label{fig:conversion} Demonstration of mode conversion
$LA\rightarrow LA+TA$  near the critical angle from Ge to Si.  (A)
The ratio of transmission converted to $LA$; (B) The ratio of
transmission to $TA$;  (C) The energy transmission $T_n(\omega)$ as
a function of angular frequency $\omega$   across the Ge-Si (100)
interface for $LA$ waves at different angles of incident from Ge-Si.
}
\end{figure}

\emph{AA conversion.} Apart from the possible AO conversion, there
are several different polarized acoustic branches. Can the the
conversion among these different acoustic branches occur at the
interface? Our simulation did not find  significant conversion among
these different polarized branches, except when the angle of
incidence is near the critical angle. The reflected and transmitted
waves are both LA waves when the wave incident is longitudinally
polarized. The same
 is true for the TA wave.  The AA conversion among different
polarized acoustic branches only takes place when the angle of
incident is close to the critical angle for wave incident from Ge to
Si. Fig.~\ref{fig:conversion} shows the transmission of the
longitudinally polarized wave incident from  Ge to Si.   The mode
conversion of $LA\rightarrow LA+TA$ is shown in pictures (A) and (B)
in Fig.~\ref{fig:conversion}. It can seen that the ratio converted
from LA to LA decreases with the increase of the angle of incident.
The ratio converted from LA to TA increases rapidly to about $0.12$
near the critical angle. This behavior of mode conversion taking
place near the critical angle maybe result from the suppression of
transmission due to the Snell law.  We also find that the presence
of the possible reflected waves influence the transmission. For
example,  in contrast to the transmission from Si to Ge, the
dependence of $LA$ mode wave transmission on frequency from Ge to Si
shows a rich character as illustrated in Fig.~\ref{fig:conversion}.
For wave with the incident angle $\theta \neq 0 $, the transmission
first decreases with the increase of frequency. But when the
frequency goes over a certain value, for example $\omega=93\, {\rm
cm}^{-1}$ for $\theta=36^{\circ}$, the transmission begins to
increase. This behavior can be understood by the mode conversion at
the interface. It can be seen from Fig.~\ref{fig:dispersion} that
the maximum frequency for $TA$ modes in Ge is about $98\, {\rm
cm}^{-1}$.  When the frequency is below this value, there are $TA$
modes for the reflected wave; but over this value, the reflected
wave cannot be converted into $TA$ modes. The transmission increases
due to lack of reflected modes, .

\subsection{Transmission across the interface between diamond and zinc-blend structure }
During the calculation of the transmission of Si-Ge interface, the
atomic masses are uniformly distributed both for Si and Ge on each
side. There is another important type of lattice structure in the
semiconductor materials. That is the zinc-blend structure, where the
atomic masses in the sublattice are different.
\begin{figure}[ht]
\includegraphics[width=0.8\columnwidth]{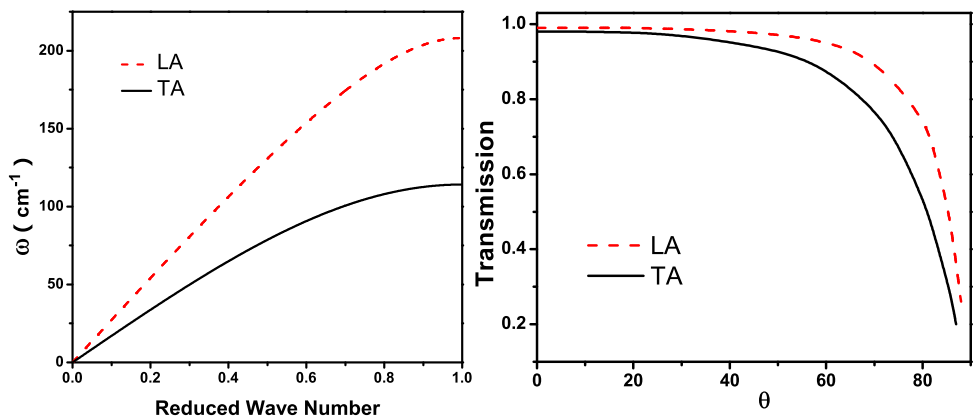}
\caption{ \label{fig:gap} Left figure: Phonon dispersion for
acoustic waves in GaP.  Right figure: Dependence of phonon
transmission for LA waves on the incident angle $\theta$ at angular
frequency $ \omega = \mbox{15 cm}^{-1} $ from Si to GaP.  }
\end{figure}
 Here we consider an
example of gallium phosphide(GaP). To simplify the problem, we only
consider the valence-force and neglect the Coulomb interaction
between the charges when computing the phonon dispersion of GaP.
This is a valid approximation when we consider the acoustic phonons
that play major role in the phonon transmission across the
interface. The calculated acoustic phonon dispersion curves along
$\Gamma -L$ for Gap is plotted in Fig.~\ref{fig:gap}. The endpoint
values of frequencies along $\Gamma -L$ for $LA$ and $TA$ are about
$208 \mbox{ cm}^{-1}$ and $116 \mbox{ cm}^{-1}$, respectively. It
can be seen that valence-force constants approximately describe the
acoustic branches \cite{gap} of GaP.  The results of phonon
transmission for $LA$ and $TA$ modes across a Si-GaP interface are
illustrated in Fig.~\ref{fig:gap}. We find a consistent dependence
on frequency, spatial angular and a similar phenomena of mode
conversion with that of Si-Ge interface. The dependence of
transmission on the angle of incidence is illustrated in
Fig.~\ref{fig:gap}.

\subsection{ Temperature dependence of Kapitza conductance  }

\begin{figure}[h]
\includegraphics[width=0.5\columnwidth]{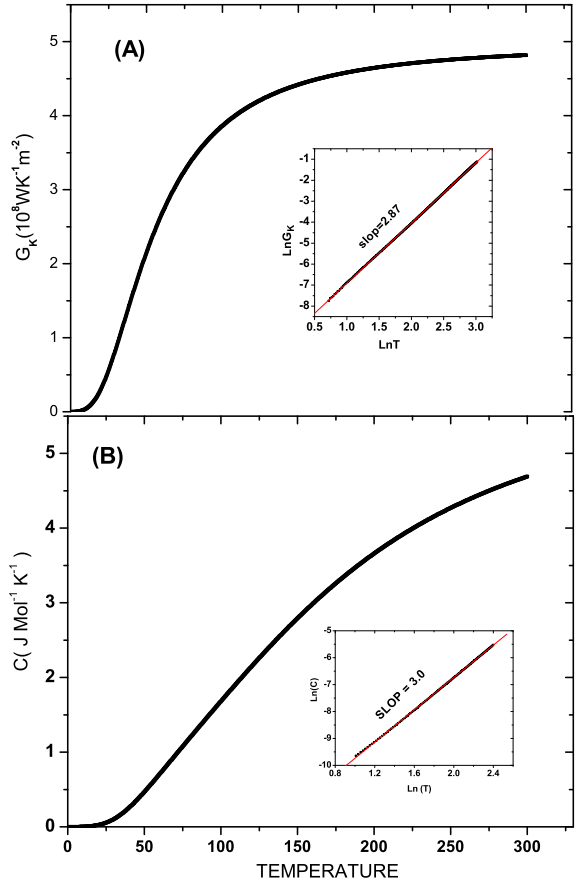}
\caption{ \label{fig:kapitza}  (A) The temperature dependence of
Kapitza conductance for Si-Ge(100) interface. (B) The corresponding
heat capacity of Si.  }
\end{figure}

The Kapitza conductance with the change of temperature is calculated
using Eq.~(\ref{current}), and are illustrated in
Fig.~\ref{fig:kapitza}(A). The Kapitza conductance for Si-Ge $[100]$
interface calculated from our model is $G_{K}=4.6\times10^{8}\, {\rm
WK}^{-1}{\rm m}^{-2}$ when $T=200\,$K. When the temperature goes
beyond $200\,$K, we find that the Kapitza conductance changes little
with the temperature and is saturated. For comparison, we plotted
the heat capacity of Si in Fig.~\ref{fig:kapitza}(B). It can be seen
that the heat capacity continues to increase with the temperature
when $T>200\,$K. In comparison with the heat capacity, the
saturation of Kapitza conductance can be accounted by negligible
contribution of energy transmission from high frequency at  low
temperatures, as illustrated in inset of Fig.~\ref{fig:kapitza}. The
Kapitza conductance scales as $T^{2.87}$, while the heat capacity
scales as $T^{3}$ in accordance with the Debye model.  We have
sampled enough points in the first Brillouin zone to ensure that
Kapitza conductance and heat capacity converge numerically.
However, due to the small deviation from the value of three, we
cannot rule out the possibility that the exponent for Kapitza
conduction is also 3. Ref.~\onlinecite{dayang} reported Kapitza
conductance scaled $T^{3}$ at  low temperature for $fcc$ interface
irrespective of the properties for the left and right lead. The
temperature dependence of Kapitza conductance is an intriguing
problem \cite{ETswartz}, though much experimental work has been done
on this field. Most experiments gave $T^{\alpha} \,\, \mbox{with}
\,\,\, \alpha\leq3$ for solid interface as reviewed in
Ref.~\onlinecite{ETswartz}. So far no experiment result is available
for the temperature dependence for the Si-Ge interface. Compared
with the results of Ref.~\onlinecite{dayang} our discrepancy from
$T^{3}$ comes from the anisotropy of the energy transmission because
of the diamond structure used for calculation of the transmission,
while Ref.~\onlinecite{dayang} took an isotropic assumption for
their calculation.

\section{A new simplified approximate formula  }
\begin{figure}[hbt]
\includegraphics[width=0.5\columnwidth]{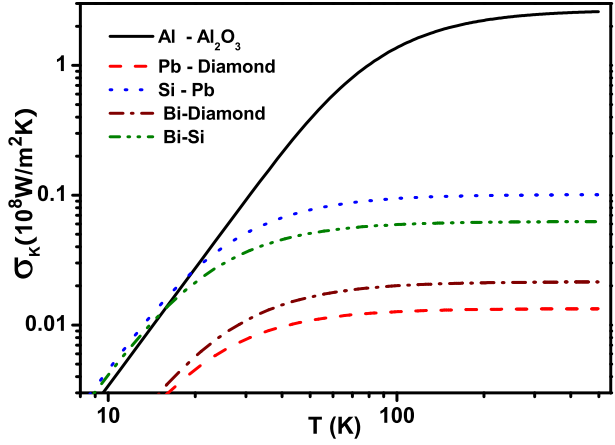}
\caption{ \label{fig7_kapitza}  The temperature dependence of
Kapitza conductance calculated from Eq.~\ref{simplify}. }
\end{figure}

To  get a simplified  expression, we make the follow approximation
on the basis of the features of transmission discussed above.  (1)
For transmission with a given incident angle $\theta$, we assume
that it is independent of the frequency.  (2) We neglect the
dependence on the azimuth angle. (3) The conductance can be
calculated from either the left or the right side.\cite{wang} They
give the same result across the interface. It is found that the
transmission from the high group velocity to the low velocity side
has a simpler feature.  We calculate the transmission from the high
to the low velocity side.  The dependence of transmission on the
incident angle for a polarized waves is formulate as
$\mathcal{T}^{n}=\mathcal{T}^{n}_0\cos \theta$. Here
$\mathcal{T}^{n}_0 $ is the transmission at $\omega =0$ with the
angle of incident $\theta =0$, and is given by
\begin{subequations}
\label{transmission}
\begin{eqnarray}
 \label{reftran}
 \mathcal{T}_{0}^{n} & = & \frac{4Z_1^nZ_2^n}{(Z_1^n+Z_2^n)^2}  \\
 \label{reftran2}
  Z_i^n = & \rho _i \frac{v^n_i}{l_i}, & \mbox{with  } i=1,2.
\end{eqnarray}
\end{subequations}
Here $Z_i^n$ is the acoustic impedance for the polarized wave and
$l_i$ is the lattice constant. We have incorporated the effect of
difference in  lattice constants of the crystal. It can be seen from
Fig.~\ref{fig:critical_angle} that the cosine function is a valid
approximation.  (4) Mode conversion at the interface can be
neglected. (5) Since thermal conductance is contributed mainly by
acoustic waves, we can assume that the upper limit of frequency of
acoustic waves is characterized by the Debye temperature of the side
with the group velocity.  A linear dependence of phonon dispersion
relation $\omega =qv_n$ is used  because thermal conductance is
mainly due to acoustic waves. The group velocity $v_n^z$ along $z$
in Eq.~\ref{current} is given by $v_n^z=v_n\cos \theta$. Under these
conditions, we have a simplified thermal conductance (Kapitza
conductance) as
\begin{equation}
 \label{simplify}
 \sigma _K=\frac{1}{12\pi ^2} \sum_n \frac{\mathcal{T}_0^{n}}{v_n^2} \cdot \frac{k_B^4T^3}{\hbar ^3} \cdot
 \int_0^{x_D}{\frac{x^4e^x}{(e^x-1)^2}dx}.
\end{equation}
Here $v_n$ is the group velocity of the incident side with the large
group velocity. The upper limit $x_{D}$ is given as
$x_{D}=\frac{\theta _D}{T}$, where $\theta _D$ is the Debye
temperature for the other side with low group velocity.
\begin{table}[ht]
 \setlength{\abovecaptionskip}{0pt}
\setlength{\belowcaptionskip}{10pt} \caption{\label{tab:parameter}
 Values for mass density $\rho$, lattice constant $l$, the longitudinal group
velocity  $v_L$, the  transversal group velocities $v_T$, heat
capacity $C_v$ and the Debye temperature $\theta _D$. These values
are from Ref.~\onlinecite{ETswartz, njp,kittel}.}
\begin{center}
\renewcommand{\arraystretch}{0.8}
\tabcolsep=4pt
\begin{tabular*}{14cm}{ccccccc}
\hline \hline
& $\rho (10^3 \mbox{Kg}/m^3 )$ & $l$(\AA)  & $v_L (\mbox{ Km}/s)$   & $v_T (\mbox{ Km}/s)$ & $C_v (10^6\mbox{J}/m^2\mbox{K})$  & $\theta _D$(K) \\
\hline Si & 2.329  & 5.43  & 8.43   & 5.84  &  1.63  &  640 \\
Al & 2.699 &  4.05 & 6.24 &  3.04 & 2.42 & 394    \\
$\mbox{Al}_2\mbox{O}_3$ & 3.97  &  4.76 & 10.89  & 6.45  &  2.656  & 1024  \\
Bi & 9.79 &   4.75 & 1.972 &  1.074 &  1.194 &  120  \\
Pb & 11.598 & 4.95 & 2.35  &  0.97  &  1.463 &  105  \\
Diamond & 3.515 &  3.57 & 17.52 & 12.82  &  1.8278 &  1860 \\
 \hline \hline
\end{tabular*}
\end{center}
\renewcommand{\arraystretch}{1.0}
\end{table}
We use Eq.~(\ref{simplify}) to calculate a few interfaces across
which the experimentally measured thermal conductance is available.
The parameters taken are shown in Table~\ref{tab:parameter}. The
results of calculation are shown in Fig.~\ref{fig7_kapitza}.

We list the experimental values and our calculated values in
Table~\ref{tab:kaptiza}. \begin{table}[bt]
 \setlength{\abovecaptionskip}{0pt}
\setlength{\belowcaptionskip}{10pt} \caption{\label{tab:kaptiza}
Comparison of thermal conductance across interface for the
experimental measured values and the calculated values at $T=200K$.
The experimental values are  from Ref.~\onlinecite{experimental}.
The values are in the units of $10^8W/m^2K$.}
\begin{center}
\tabcolsep=5pt
\renewcommand{\arraystretch}{0.7}
\begin{tabular*}{14cm}{cccccc}
\hline \hline
& Al/$\mbox{Al}_2\mbox{O}_3$ & Pb/Diamond &  Bi/Diamond & Pb/Si & Bi/Si \\
\hline
Exp. Values & 1.2 & 0.125 & 0.06 & 0.13 & 0.1025 \\
Cal. Values & 2.5 & 0.013 & 0.02 & 0.10 & 0.0640 \\
 \hline \hline
\end{tabular*}
\end{center}
\renewcommand{\arraystretch}{1.0}
\end{table} It can be seen that the calculated values
are consistent with the experimental values except for the interface
Pb/Diamond. The large deviation for Pb/Diamond interface between the
calculated and experimental values maybe come from the nonlinear
scattering effect at the interface. It cannot be explained in the
reach of the method employed in this paper.

\section{discussion and Conclusion }
  We have studied the features of phonon transmission across the epitaxial
  interfaces by lattice dynamic approach. The  transmission is found
  to change slightly with the frequency for polarized waves with a
  given incident angle from the high group velocity side to the
  low group velocity side. The dependence of transmission on the
  azimuth angle is related to the symmetrical properties of the
  interface for waves with larger angle of incident. A critical
  angle exists for transmissions from
 the low group velocity side to the higher group velocity side.
 Dependence of transmission on the mode conversion at the interface is trivial
 except when the incident angle is close to the critical angle.  Thermal conductance across
 the epitaxial  interfaces is dominantly contributed by the acoustic
 waves and is insensitive to the mass distribution of different
 lattice positions. Kapitza conductance across the Si-Ge(100)
 interface shows a $T^{2.87}$ dependence on temperature, which is
 very close to $T^3$ behavior.  A simplified formula for the estimation of thermal
 conductance across the interface is proposed in the light of
 features found by lattice dynamic approach.  We remark that this formula
 can give a valid estimation of Kapitza conductance across solid
 epitaxial interfaces when nonlinear phonon scattering is
 unimportant.  We think nonlinear scattering sometime maybe important for the
 interfaces between materials with the higher Debye temperature and
 the lower Debye temperature, such as Pb-diamond.  Such a nonlinear effect
 cannot be captured in the current lattice dynamic approach.  From
 the calculation, the nonlinear effect maybe contribute to the
 increase of thermal conductance across the solid interface by
 breaking the selecting rules between the linear phonon dispersion
 relations. The phase relations among the incident wave and the reflected, transmitted waves have not been
 considered  because they will not influence the energy transmission across the interface.
  A possible existence of lattice dislocation or disorder
 is not included in the present lattice dynamic approach.

\section*{ACKNOWLEDGEMENTS}
We think Dr. Jingtao L\"u for careful reading of the manuscript.
This work is supported in part by a Faculty Research Grant of
National University of Singapore.


\begin{thebibliography}{01}

\bibitem{dgcahill} D. G. Cahill \textsl{et al.},
J.  Appl. Phys. \textbf{93}, 793 (2003).

\bibitem{ETswartz} E. T. Swartz and R. O. Pohl,
Rev. Mod. Phys. \textbf{61}, 605 (1989).

\bibitem{Kapitza}  P. L. Kapitza, J. Phys.
(Moscow) \textbf{4}, 181 (1941).

\bibitem{walittle} W. A. Little,
Can. J. Phys.   \textbf{37}, 334 (1959).

\bibitem{G-Chen} G. Chen, Phys. Rev. B \textbf{57}, 14958
(1998).

\bibitem{experimental} R. M. Costescu, M. A. Wall, and D. G. Cahill,
Phys. Rev. B  \textbf{67}, 054302 (2003); H.-K. Lyeo and D. G.
Cahill,  Phys. Rev. B  \textbf{73}, 144301 (2006).


\bibitem{dayang} D. A. Young and H. J. Maris,
Phys. Rev. B  \textbf{40}, 3685 (1989); R. J. Stoner and H. J.
Maris, Phys. Rev. B  \textbf{48}, 16373 (1993).

\bibitem{wang} J. Wang and J.-S. Wang, Phys. Rev. B \textbf{74}, 054303 (2006).

\bibitem{zhao}  H. Zhao and J.
B. Freund,  J. Appl. Phys. \textbf{97}, 024903 (2005); ibid,
\textbf{97}, 109901 (2005).


\bibitem{numericalrecipes} W. H. Press, {\textit{et al.}}, \textsl{Numerical Recipes in C},
2nd Ed.,  Cambridge Univ.  Press, 2002, p.~59.

\bibitem{dwbrenner} D. W. Brenner \textit{et al.},
J. Phys.: Condens. Matter. \textbf{14}, 783 (2002);  J. Tersoff,
Phys. Rev. B \textbf{39}, 5566 (1989).

\bibitem{giannozzi}R. Tubino, L. Piseri, and  G. Zerbi, the J. Chem. Phys. \textbf{56}, 1022(1972);
 Giannozzi  \textsl{et al.}, Phy. Rev. B  \textbf{43}, 7231 (1991).

 \bibitem{abinit} K. Parlinski, Z. Q. Li, and Y. Kawazoe,
Phys. Rev. Lett.  \textbf{78}, 4063 (1997).

\bibitem{auld} B. A. Auld, \textsl{Acoustic Fields and Waves in Solids Volume II},
2nd Ed.,  Robert E. Krieger Publishing Company, Inc 1990, p.~22.

\bibitem{gap} R. Banerjee and Y. P. Varshni, J. of Phys. Soc. Japan \textbf{30}, 1015(1971); E. O. Kane,  Phys. Rev. B \textbf{31}, 7865(1985).

\bibitem{njp} B. Krenzer, A. Janzen, P. Zhou, D. Linde and M.H. Hoegen  New J. Phys\textbf{8},
190(2006).

\bibitem{kittel} C. Kittel, {\textit{et al.}}, \textsl{Introduction to Solid State Physics},
7th Ed.,  John Wiley  Sons Inc, 1996.


\end{thebibliography}
\end{document}